\documentclass[conference]{IEEEtran}
\makeatletter
\newcommand*\titleheader[1]{\gdef\@titleheader{#1}}
\AtBeginDocument{%
  \let\st@red@title\@title
  \def\@title{%
    \bgroup\normalfont\large\centering\@titleheader\par\egroup
    \vskip1.5em\st@red@title}
}
\makeatother

\usepackage{cite}
\ifCLASSINFOpdf
  \usepackage[pdftex]{graphicx}

\else

  \usepackage[dvips]{graphicx}

\fi

\usepackage{subcaption}
\usepackage[hyphens]{url}
\usepackage[font=footnotesize]{caption}
\usepackage{xcolor}
\usepackage{algorithm}
\usepackage{algpseudocode}
\usepackage{ifthen}
\usepackage{ulem}
\usepackage{wrapfig}
\usepackage{epstopdf}

\newcommand{\version}{paper}
\newcommand{\cond}[2]{\ifthenelse{\equal{\version}{#1}}{#2}{}}
\makeatletter
\renewcommand{\ALG@beginalgorithmic}{\small}
\makeatother
\newcommand{\myfigure}[1]{Fig.~\ref{#1}}

\newcommand{\info}[1]{\cond{info}{\textcolor{blue}{#1}}}
\newcommand{\depecated}[1]{\cond{info}{\textcolor{gray}{#1}}}
\newcommand{\todo}[1]{\cond{info}{\textit{\textcolor{red}{TODO: #1}}}}

\renewcommand{\emph}[1]{\textit{#1}}
\titleheader{2016 IEEE 24th International Requirements Engineering Conference}
\titleheader{This paper has been accepted in IEEE Journal on Selected Areas in Communications,\\ vol. 35, no. 11, pp. 2468-2478, Nov. 2017. (doi: 10.1109/JSAC.2017.2760418)}

\title{NFV and SDN - Key Technology\\ Enablers for 5G Networks}

\author{
	\IEEEauthorblockN{Faqir Zarrar Yousaf, Michael Bredel, Sibylle Schaller, and Fabian Schneider}
	\IEEEauthorblockA{
		NEC Laboratories Europe\\
		69115 Heidelberg, Germany\\
		Email: \{zarrar.yousaf, michael.bredel, sibylle.schaller, fabian.schneider\}@neclab.eu
	}
}
\begin{document}
\maketitle

\begin{abstract}
Communication networks are undergoing their next evolutionary step towards 5G. The 5G networks are envisioned to provide a flexible, scalable, agile and programmable network platform over which different services with varying requirements can be deployed and managed within strict performance bounds. In order to address these challenges a paradigm shift is taking place in the technologies that drive the networks, and thus their architecture. Innovative concepts and techniques are being developed to power the next generation mobile networks. At the heart of this development lie \textit{Network Function Virtualization} and \textit{Software Defined Networking} technologies, which are now recognized as being two of the key technology enablers for realizing 5G networks, and which have introduced a major change in the way network services are deployed and operated. For interested readers that are new to the field of SDN and NFV this paper provides an overview of both these technologies with reference to the 5G networks. Most importantly it describes how the two technologies complement each other and how they are expected to drive the networks of near future. 

\bigskip
\end{abstract}


%
\IEEEpeerreviewmaketitle


%

\section{Introduction}
\label{sec:introduction}
Communication networks have evolved through three major generational leaps following the technology trends and constantly evolving user demands. The first evolutionary jump was from the first generation, known as 1G, to the second generation, i.e. 2G, when the mobile \emph{voice} network was digitized. The next evolutionary jump from 2G to 3G was made in order to fulfill the users' ever increasing demand for data and service quality. The proliferation of sophisticated user platforms, such as smart phones and tablets, and mushrooming new bandwidth intensive mobile applications further fueled user appetite for bandwidth and quality. This has led to the next evolutionary leap towards 4G, which has made mobile networks provide a true \emph{wireless broadband service} to its customers. With the enhanced options offered by 4G, new use cases, such as in health, automotive, entertainment, industrial, social, environmental etc. sectors, with diverse service requirements have been introduced. Services are innovating rapidly with exceeding reliance on the mobile network infrastructure for their connectivity needs. With such evolution and the Internet transforming towards an Internet-of-Things, the notion of a customer has changed from human customers only to now also include cars, sensors, consumer electronic items, energy meters etc. With such a diverse customer base, the mobile network not only has to manage the burgeoning data volume, but at the same time ensure that customer service requests are being adequately fulfilled by the network, meeting the respective quality-of-service or quality-of-experience requirements. In order to meet the data and service requirements, the network operators are constantly expanding and upgrading their network infrastructure, resulting in increased capital and operational expenditures (capex and opex). However, in view of the intense competition and falling prices, the average revenue per user is not increasing proportionately resulting in lower return on investment. Thus, in order to reduce costs and increase revenue mobile networks need to take their next evolutionary leap towards 5G, which now not only addresses the mobile edge, but also the core network.

\begin{figure}[!h]
\includegraphics[trim={1cm 1cm 1.5cm 1.3cm}, width=0.48\textwidth]{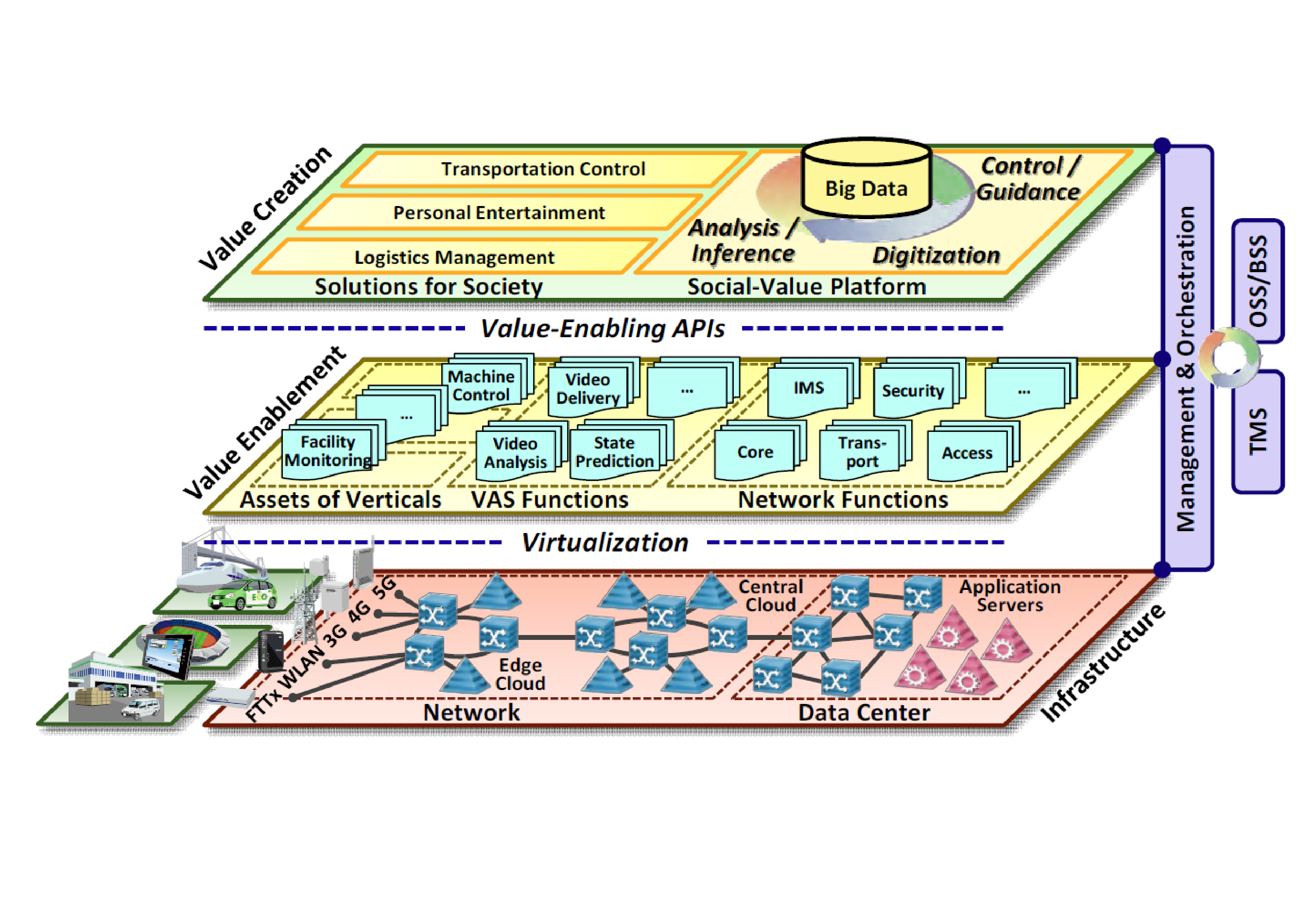}
 \caption{A 5G System Vision \cite{necVision}.}
 \label{fig:5g_vision}
 \end{figure}

\subsection{5th Generation Networks - Vision}
\label{sec:5gVision}
5G networks, also referred to as beyond 2020 communications systems, represent the next major phase of the telecom industry. The three main features that shall characterize a 5G network will be its ability to support \emph{Enhanced Mobile Broadband, Massive Machine Type Communication} and the provisioning of \emph{Ultra-reliable Low Latency Communication} services. This entails 5G networks to provide increased peak bit rates at Gbps per user, have higher spectrum efficiency, better coverage, and support for a massively increased number of diverse connectable devices. In addition, 5G systems are required to be cost efficient, reliable, flexibly deployable, elastic, agile, and above all programmable. These are ambitious and highly challenging requirements that have implications on both the mobile radio access network as well as the mobile core network, and thus require a major re-design and re-engineering of both the architecture and the technologies. In view of these ambitious requirements, new innovative methods and systems are being explored and evaluated in order to meet the challenging performance goals of 5G networks. 

First commercial deployments of 5G networks are expected in 2020. Different stake-holders have expressed their respective vision of a 5G network, and \myfigure{fig:5g_vision} illustrates one such high-level vision~\cite{necVision} where the 5G network eco-system is depicted as a three-tier model. Shown at the lowest level are physical resources and assets such as compute, network, storage, which are distributed and available in the back-end data centers, core network infrastructures, and radio access networks. These physical resources are abstracted to create a virtualized second level where network functions and other value-added application functions are enabled as virtualized instances or entities. The top-level consists of heterogeneous services that shall consume the APIs exposed by the virtualized entities below in order for them to provide their respective services transparently and in isolation to each other over a common network platform while meeting their respective operational and functional service requirements. 

\subsection{5G Slicing Concept \& Challenges}
\label{sec:5gSlicing}
The vision of 5G networks discussed above leads to a very important concept of \textit{slicing} that has become a central theme in 5G networks. Network slicing allows network operators to open their physical network infrastructure platform to the concurrent deployment of multiple logical self-contained networks, orchestrated in different ways according to their specific service requirements; such network slices are then (temporarily) owned by tenants. As these tenants have control over multiple layers, i.e. the physical layer, the virtualization layer, and the service layer, of a 5G infrastructure, they are also called \textit{verticals}: That is, they integrate the 5G infrastructure vertically. The availability of this vertical market multiplies the monetization opportunities of the network infrastructure as (i) new players, such as automotive industry and e-health, may come into play, and (ii) a higher infrastructure capacity utilization can be achieved by admitting network slice requests and exploiting multiplexing gains. With network slicing, different services, such as, automotive, mobile broadband or haptic Internet, can be provided by different network slice instances. Each of these instances consists of a set of virtual network functions that run on the same infrastructure with a tailored orchestration. In this way, very heterogeneous requirements can be provided on the same infrastructure, as different network slice instances can be orchestrated and configured separately according to their specific requirements, e.g. in terms of network quality-of-service. Additionally, this is performed in a cost efficient manner as the different network slice tenants share the same physical infrastructure.

\begin{figure}[t!]
	\centering
	\includegraphics[width=\linewidth]{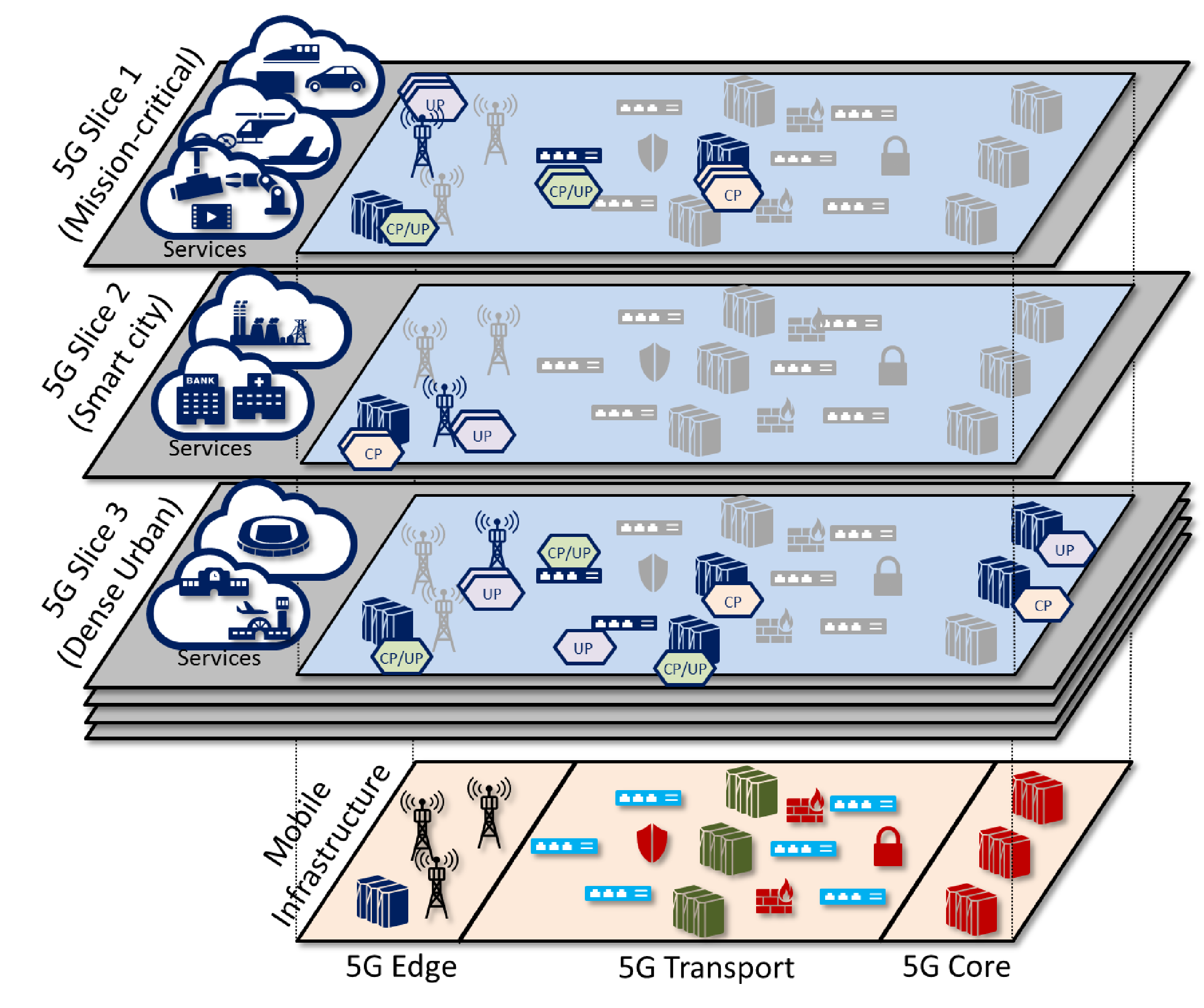}
	\caption{Network Slicing in 5G as envisioned by the NGMN project.}
	\label{fig:5gslice}
\end{figure}

While the network slicing concept has been proposed recently~\cite{ngmnNS}, it has already attracted substantial attention and several standardization bodies started working on it. 3GPP has is working on the definition of requirements for network slicing~\cite{TR23799}, whereas NGMN identified network sharing among slices as one of the key 5G issues~\cite{ngmn5G}. A Network Slice is defined by NGMN as ’a set of network functions, and resources to run these network functions, forming a complete instantiated logical network to meet certain network characteristics required by the service instance(s)’. According to NGMN, the concept of network slicing involves three layers namely (i) service instance layer, (ii) network slice instance layer, and (iii) resource layer. The service instance layer represents the end-user and/or business services, provided by the operator or the 3rd party service providers, which are supported by the network slice instance layer. The network slice instance layer is in turn supported by the resource layer, which may consist of physical resources such as compute, network, memory, storage etc, or it may be more comprehensive as being a network infrastructure, or it may be more complex as network functions. \myfigure{fig:5gslice} depicts this concept where the resources at the resource layers are dimensioned to create several subnetwork instances, and network slice instances are formed that may use none, one or multiple sub-network instances. 

The 5G mobile network system is thus going to be multi-tiered and slices need to be deployed and managed at each level resulting in not only a complex architecture, but posing enormous challenges in terms of 5G network sliced infrastructure and traffic management. In this regard some of the principal key are:

\begin{enumerate}
	\item Seamless and flexible management of physical and virtualized resources across the three tiers.
	\item Agile end-to-end service orchestration for each respective service vertical, where each vertical may have multiple service instances.
	\item Enabling end-to-end connectivity services to each service instance, which is also programmable.
\end{enumerate}

In consideration of the above challenges, two key technologies are being developed in order to cater scalability, flexibility, agility, and programming requirements of 5G mobile networks: \emph{Network Function Virtualization (NFV)} and \emph{Software Defined Networking (SDN)}. The inherent potential and recent advances in the area of NFV and SDN have made them being recognized as key technological enablers for the realization of a carrier cloud, which is a key component of the 5G system. NFV is being designed and developed specifically in terms of addressing flexibility, agility and scalability requirements, and it leverages on the recent advances in cloud computing and their support for virtualized services. On the other hand, SDN is being developed in order to make the connectivity services provided by 5G networks programmable, where traffic flows can be dynamically steered and managed in order to gain maximum performance benefits. 
%
%
However, there are numerous challenges for making SDN and NFV deployable and carrier-grade \cite{carrierChallenges}. Consequently, the Open Networking Foundation (ONF) and the ETSI NFV Industry Special Group have been formed to standardize various aspects of SDN and NFV-enabled networks respectively. 

The subsequent sections provide an overview of the main technological and architectural features of NFV and SDN, and describe how these two technologies can realize a 5G core network. A detailed discussion on how NFV and SDN complement each other towards realizing a functional 5G core network is also provided.
\section{NFV and MANO Systems}
\label{sec:mano_systems}
%

Obviously, the complex architecture of upcoming 5G networks calls for an efficient management framework that provides a uniform and coherent orchestration of various resources across the multiple layers of the 5G ecosystems. Network Function Virtualization and their Management and Orchestration (MANO) systems offer themselves as elegant solutions, aiming at decreasing cost and complexity of implementing new services, maintaining running services, and managing available resources in existing infrastructure. Thus, in the following we provide a detailed introduction to NFV and MANO systems and give an overview of various open-source projects and solutions available today.

\subsection{Network Function Virtualization}
\label{sec:mano_systems:nfv}

\info{
\begin{itemize}
	\item \sout{Motivate the need for NFV}
	\item \sout{Explain the terms: NFV, VNF, SFC, and Network Service}
	\item Explain the "security" model in the sense that if a service goes down 
	it will just restarted
	\item Explain horizontal (and vertical) scaling. Mention why scaling is 
	challenging
\end{itemize}
\medskip
}

The rise of powerful general-purpose hardware, cloud computing technology, and flexible software defined networks, led to the first idea of virtualizing classical network functions, such as routers, firewalls, and evolved packet cores. These network functions, which have been executed on dedicated and often specialized hardware before, now run as software applications in virtual machines on top of cloud infrastructure. Thus, the operation of dedicated network middle-boxes transfers into the operation of virtual machines and software, which paves the way to reduce capital expenditures by using common-of-the-shelf hardware and to apply existing management practices and tools from the cloud computing space in order to automate network operation tasks and reduce operational expenditures. Hope is that NFV and networked systems benefit from automation and unified ecosystems the same way cloud environments did already. Moreover, NFV systems could embrace the high-availability model of cloud systems. Rather than trying to build an architecture that can't fail, which is the dominant approach in today's telco world, NVF aims at creating an architecture that builds failure management into every part of the system and horizontally partitions it to limit single points of failure.

The first generation of NFV system implementations transferred existing monolithic applications to big virtual machine appliances, each representing a single \textit{Virtual Network Function} (VNF). Multiple VNFs are then chained together using a \textit{Service Function Chain}, which determines how packets are forwarded from one VNF to another, to constitute a \textit{Network Service}. This already improved flexibility as well as manageability of networks, as operators can use existing cloud management tools, such as Puppet, Chef, and JuJu. But at the same time, it also allowed operators to use existing and well-known paradigms of traditional networks, like high-availability concepts using redundant systems and hot-standbys.

However, it has been reported, e.g. in~\cite{taylor:2015:01}, that the model of using fat virtual machines and traditional high-availability and performance concepts does not translate well to the cloud. Simple ports of software, which was originally designed to run on specialized hardware appliances, are often not able to deliver performance and high-availability on standard cloud environments. For instance, cloud systems, hypervisors, and virtual machines introduce an overhead in input/output operations, which limits the performance of packet processing significantly. In addition, these legacy solutions often lack mechanisms to scale horizontally, i.e. to add more nodes to (or remove nodes from) a system in order to meet performance requirements. Moreover, solutions that strive to avoid failure by vertical integration of failure management to an underlying high availability platform, often fail to adapt to the cloud-native high-availability paradigm, where service instances can be killed and restarted any time. This is because underlying assumptions and mechanisms are very different~\cite{nadareishvili:2012:01}.


Today's approaches, therefore, move even further and aim at a more cloud-native software design for network applications with a much smaller footprint; not running in fat virtual machines but in slim container solutions. This however, imposes even more challenges on the NFV management as the number of NFV entities, which need to be orchestrated, increases significantly. Thus, we elaborate on management and orchestration systems in the following.

\subsection{Management and Orchestration}
\label{sec:mano_systems:mano}
\info{
\begin{itemize}
	\item \sout{In the context of ETSI MANO, explain the terms: VNFC, VDU, VIM, NFVI, VNFM, NFVO, EMS, OSS/BSS}
	\item Elaborate on MANO systems in general. Relate to Cloud Computing 
	"MANO" systems, like JuJu
	\item \sout{Elaborate on the ETSI reference model}
	\item Explain current (open-source) solutions, like OSM, ECOMP+OpenO=ONAP, 
	OpenBaton, Tacker, SONATA
	\item Elaborate on SONATA and its DevOps approach.
\end{itemize}
\medskip
}


In general, NVF Management and Orchestration systems aim at a simplified handling of complex network services using NFV technology. To this end, MANO systems have to manage virtualized infrastructure, such as cloud systems, communication and network infrastructure, like Software Defined Networks, NFV entities, like Virtualized Network Functions, and the various life-cycles of all these components. Virtualized Network Functions are often implemented as virtual machine or container images. In view of the multi-tiered architecture vision for 5G and the related slice concept discussed earlier, a 5G network is mainly composed of three layers, namely the resource layer, the network slice instance layer and the service instance layer. Each of these respective layers needs to be managed in coordination with other layers. 
How these management plane entities manage and orchestrate between physical or virtual resources at their respective planes, and more importantly, how they coordinate with each to deliver an effective 5G mobile network service platform is indeed a challenging proposition and mandates the design and development of an effective NFV Management and Orchestration system that is sensitive to the stringent carrier requirements. 

\begin{figure}[t]
 \includegraphics[trim={0cm 0cm 1.5cm 0cm}, width=0.465\textwidth]{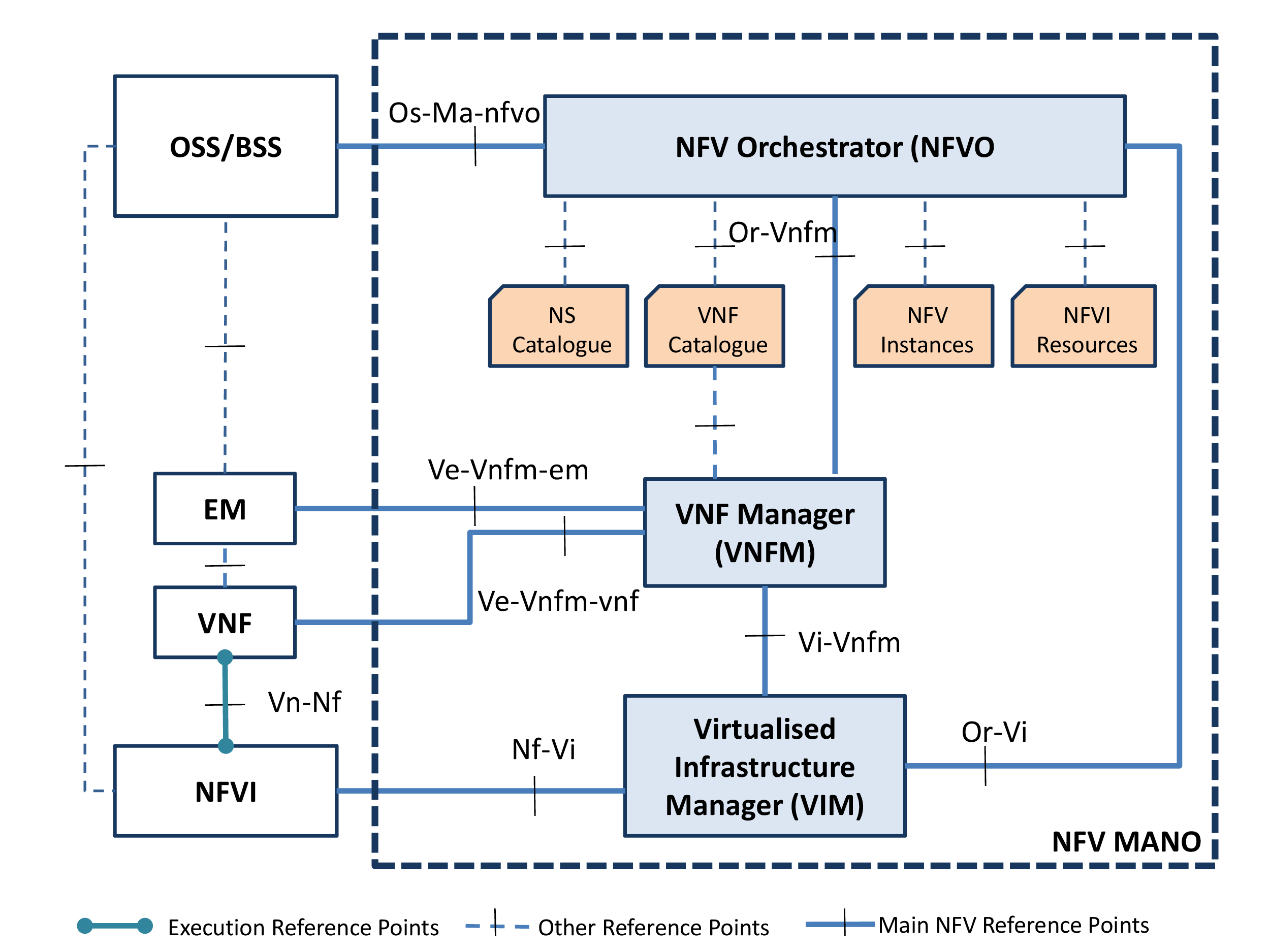}
 \caption{The NFV Management and Orchestration (MANO) framework as specified by 
 ETSI~\cite{etsi-mano}. The figure depicts the various components and reference points. It 
 clearly shows the three layers of NFV orchestration, VNF management, and 
 infrastructure management.}
 \label{fig:mano_systems:etsi_mano}
\end{figure}
\subsubsection*{\textbf{ETSI MANO Framework}}
\label{sec:mano_systems:mano:etsi}
The most relevant NFV MANO framework today is the reference model specified by ETSI and depicted in \myfigure{fig:mano_systems:etsi_mano}. This framework has three main functional blocks namely the \textit{Virtualized Infrastructure Manager} (VIM), the\textit{ Virtual Network Function Manager} (VNFM), and the \textit{Network Function Virtualization Orchestrator} (NFVO).

The NFVO manages network services. Thus, it is responsible for the on-boarding process of network service descriptions, which specify network services, and the overall life-cycle management of network services as such. It ensures end-to-end service integrity that is formed by multiple VNFs interconnected by virtual links. Therefor, the NFV Orchestrator offers reference points to external systems and might be connected to  legacy Operating Support Systems (OSS) and Business Support Systems (BSS). Moreover, the NFV Orchestrator is connected to additional data repositories, such as the network service catalog, the VNF catalog, the instance catalog, and an NFV Infrastructure resource database, which contain relevant information about the respective entities.

The VNF Manager is responsible for the life-cycle of single virtual network functions that constitute a network service. To this end, the VNF Manager might be connected to \textit{Element Managers} and Virtual Network Functions directly in order to perform actions, such as starting, scaling, and configuring the related entities. Again, external reference points allow for connections to legacy systems and facilitate a unified management and orchestration of NFV systems. Complex life-cycle events, potentially spanning multiple virtual machines, are automated and often described by Domain Specific Languages like JuJu, Puppet, and Ansible, which are executed, e.g., by process management systems.

The Virtualized Infrastructure Manager connects to NFV Infrastructures, like OpenStack cloud systems, and manages virtual network functions at the level of virtual machines and containers. Moreover, the VIM is responsible for providing connectivity between the various VNFs of a network service. Thus, it sets up the virtual links within the cloud infrastructures, e.g., by using Software Defined Networks.

In addition to the MANO framework as such, ETSI specifies various descriptors to provide metadata, such as life-cycle and monitoring information, needed to execute virtual network services and functions. To this end, the Network Service Descriptor (NSD) provides a high-level description of a network service, including all the constituent VNFs and the life-cycle events of a network service which can be interpreted and executed by the NFV Orchestrator. Likewise, VNF Descriptor (VNFD) describes a virtual network function. In addition to life-cycle events, the VNF Descriptor also includes specific information of the Virtual Deployment Units, i.e., virtual machine images or containers, and how they should be executed. For example, the VNF Descriptor describes minimal CPU requirements that must be met in order to run a certain VNF. Finally, the Network Service Descriptor, the VNF Descriptor, and other artifacts, like virtual machine images, can be combined in a \textit{Service Package} that acts as a vehicle to ship and on-board network services at a MANO service platform.

For more details on the ETSI NFV MANO framework we refer to its specification~\cite{etsi-mano} and~\cite{etsi-architectural-framework}. The related reference points are undergoing specifications at the time of writing.

\subsection{MANO Implementations}
\label{sec:mano_systems:mano_implementation}
Several open-source and commercial projects aim at implementing a MANO framework, often based on ETSI specifications as described above. Most of these projects, however, are still in an early stage but already demonstrate the abilities and advantages of a holistic service management and orchestration for network function virtualization. Below we provide a brief overview of the most relevant projects in the field.

\subsubsection*{\textbf{OSM - Open-Source MANO}}
\label{sec:mano_systems:mano:osm}
Open-Source MANO OSM~\cite{osm.webpage} is an ETSI project aiming at a reference implementation of the ETSI MANO specification. Thus, it is an operator-driven initiative to meet the requirements for orchestration of production NFV networks. OSM is based on three main software components, namely a VIM connector, Canonical JuJu, and Rift.io's Rift.ware, that reflect the three layers, i.e., Virtual Infrastructure Management, VNF Management, and NFV Orchestration layer, of the ETSI MANO framework. The OSM Virtual Infrastructure Manager connector supports multiple VIMs and natively uses OpenVIM and VMware Cloud Directory as Virtual Infrastructure Manager. JuJu Charms are used to incorporate domain knowledge on how to manage the life-cycle of virtual machines and services. Rift.io's contribution to OSM includes the NFV Orchestrator, which performs end-to-end network service delivery and drives the coherent service delivery through the resource orchestrating VIM layer and VNF configuration components in JuJu. OSM is under heavy development and Release~2 is expected to be published in early summer 2017. 

\subsubsection*{\textbf{ONAP - Open Network Automation Platform}}
\label{sec:mano_systems:mano:onap}
The ONAP project~\cite{onap.webpage} evolved from the former Open-O and ECOMP MANO projects that have originally initiated by industry. It is governed by the Linux Foundation. It is the newest player on stage and aims at building a comprehensive framework for real-time, policy-driven software automation of virtual network functions. The code is still under heavy development at the time of writing and only available to ONAP community members.

\subsubsection*{\textbf{OpenStack Tacker}}
\label{sec:mano_systems:mano:tacker}
OpenStack Tacker~\cite{Tacker.webpage} is under the big tent of OpenStack projects and aims at building an open orchestrator with a general purpose VNF Manager to deploy and operate virtual network functions on an NFV platform. It is based on the ETSI MANO architectural framework and provides a full functional stack to orchestrate VNFs end-to-end. Today, Tacker offers features like a VNF catalog, a basic VNF life-cycle management, VNF configuration management framework, and a VNF health monitoring framework. The VNF catalog makes use of the Topology and Orchestration Specification for Cloud Applications (TOSCA) language for VNF meta-data definition and OpenStack Glance to store and manage the VNF images. The Tacker VNF life-cycle management takes care of instantiation and termination of virtual machines, self-healing and auto-scaling, and VNF image updates. It also takes care of interfaces to vendor specific element management systems. Like the VNF catalog, the basic VNF life-cycle management relies on existing OpenStack services and uses OpenStack Heat to start and stop virtual machines that contain the VNF. Thus, the TOSCA templates are automatically translated to OpenStack Heat templates. OpenStack Tacker is under heavy development. At the time of writing, several crucial features, such as service function chaining and VNF decomposition, are still under discussion. \todo{@Michael: Add some references.} \todo{@Michael: Adapt and rephrase text.}

\subsubsection*{\textbf{OpenBaton}}
\label{sec:mano_systems:mano:openbaton}
OpenBaton~\cite{openbaton.webpage} is an open source project by Fraunhofer FOKUS that provides an implementation of the ETSI Management and Orchestration specification. Its main components are a Network Function Virtualization Orchestrator, a generic Virtual Network Function Manager that manages VNF life-cycles based on he VNF description, and an SDK comprising a set of libraries that could be used for building a specific VNF Manager. 

The NFV Orchestrator, which is the main component of OpenBaton, is written in Java using the spring.io framework. To interconnect the NFV Orchestrator to different VNF Managers, OpenBaton relies on the Java Messaging System. The NFV Orchestrator is currently using OpenStack as integrated Virtual Infrastructure Manager, supporting dynamic registration of NFV points of presence and deploys in parallel multiple slices consisting of one or multiple VNFs. Through this functionality the orchestrator provides a multi-tenant environment distributed on top of multiple cloud instances.

\subsubsection*{\textbf{SONATA - Agile Service Development and Orchestration in 5G Virtualized Networks}}
\label{sec:mano_systems:mano:sonata}
The SONATA open-source project~\cite{sonata.webpage} builds a service programming and orchestration framework that provides a development toolchain and a service development kit for virtualized services which is fully integrated with a service platform and orchestration system. 

To this end, the SDK component supports service developers with both a programming model and a set of software tools. It allows developers to define complex services consisting of multiple VNFs. Moreover, SONATA offers a MANO emulator such that a developer can test services in complex scenarios on a single computer without the need of a full-fletched Virtual Infrastructure Manager installation, like OpenStack. Once tested, a service provider, which can also be the service developer, can then deploy and manage the created services on one or more SONATA service platforms. Moreover, services and their components can be published in a way that they can be re-used by other developers. Thus, SONATA paves the way towards an integrated DevOps approach for network services. 

The SONATA Service Platform, which, unlike many other MANO systems, is implemented in a modular micro-service oriented way, is also based on the ETSI MANO specification. Due to the micro-service design, however, it is very flexible and a service platform operator can modify the platform, e.g., to support a desired business model, by replacing components of the loosely coupled MANO framework like plugins. Similar to OSM, the service platform today supports multiple VIMs using a Virtual Infrastructure Abstraction. Natively supported is OpenStack. Docker support is currently under development. The VNF life-cycle management, i.e. the VNF Manager in the ETSI MANO framework, is handled either by the generic Life-Cycle Manager or by Function Specific Managers that ship with any VNF. Likewise, the service life-cycle, i.e. NFV Orchestrator functionality, is managed by Service Specific Managers that come with every service. This allows to customize management and orchestration of each and every network service in a very flexible way.

\info{
	\begin{itemize}
		\item DevOps
		\item SDK and Service Platform
		\item Customizable MANO Framework
		\item Microservice approach
	\end{itemize}
}

\bigskip

\depecated{
	There are additional data repositories that may contain necessary information about network service (NS), VNF, NFV and NFVI that will enable the NFVO to perform its tasks. The MANO architecture also defines reference points for interfacing the MANO system with external entities like NFVI, OSS/BSS, VNFs and Element Managers (EM) for delivering a unified management and orchestration of a VNF system. It should be noted that the VIM manages the physical/virtualized resources of the NFVI, whereas VNFM is responsible for the Life Cycle Management (LCM) of the individual VNF(s) and there could be multiple instances of VNFM. NFVO, on the other hand, manages the Network Service (NS) by performing service orchestration and resource orchestration in order to ensure e2e service integrity that is formed by one or more VNFs interconnected over virtual links. The detailed specification of the NFV MANO framework is available in~\cite{etsi-mano, etsi-architectural-framework} whereas its reference points are undergoing specifications at the time of writing.
}

\subsection{Management and Orchestration of 5G Slices}
\label{sec:mano_systems:silces}

\begin{figure}[t]
	\includegraphics[trim={0cm 0cm 1.5cm 0cm}, 
	width=0.45\textwidth]{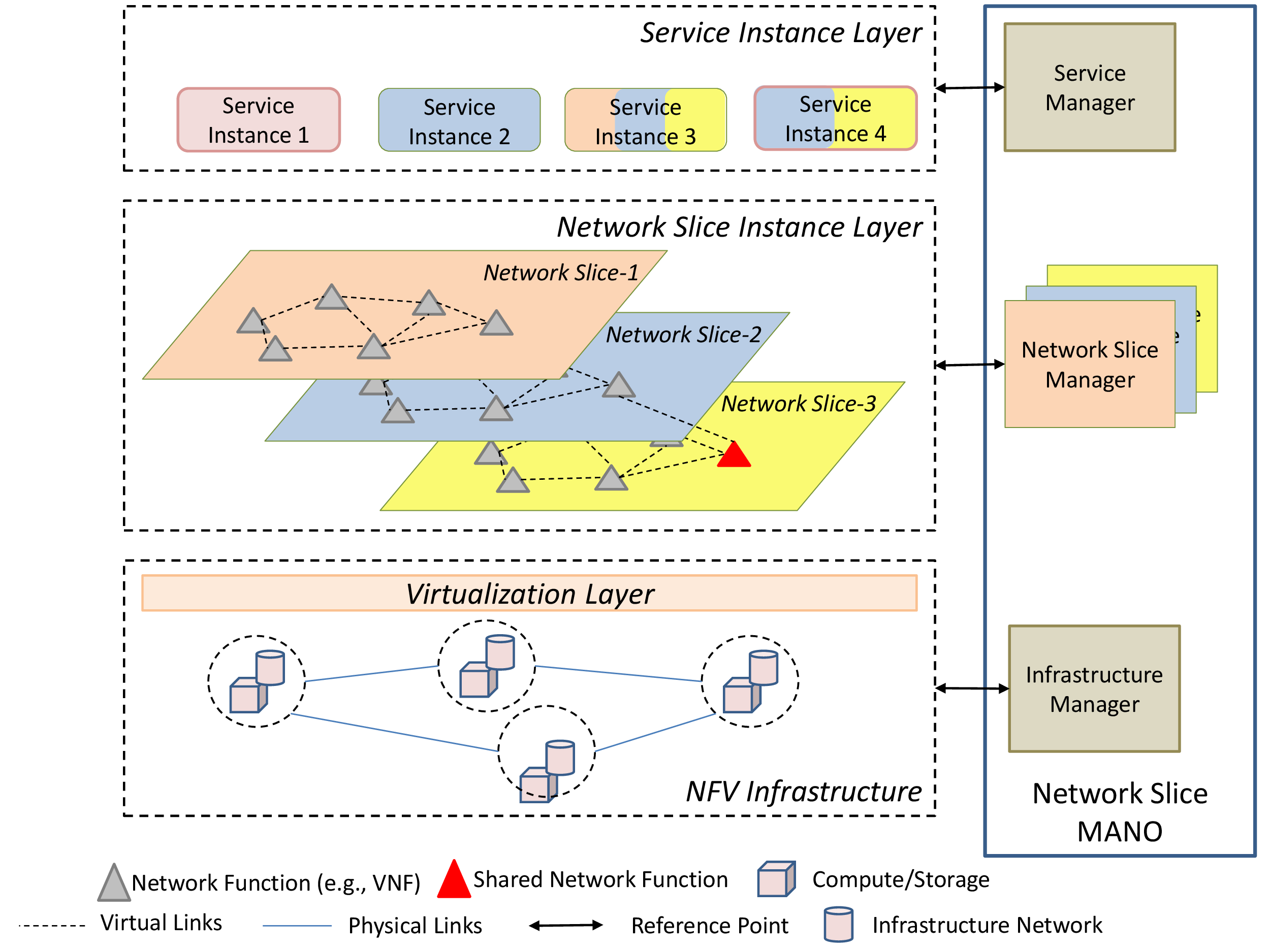}
	\caption{Network Slice Management and Orchestration (MANO) Overview.}
	\label{fig:mano_systems:slice_mano}
\end{figure}

When NFV MANO is compared to the idea of slicing in 5G networks as depicted in \myfigure{fig:mano_systems:slice_mano}, the VIM corresponds to the Infrastructure Manager, the VNFM corresponds to the Network Slice Manager while the NFVO corresponds to the Service Instance Layer. It can thus be inferred that the ETSI NFV MANO system has the required building blocks for providing a MANO framework for the 5G network slices.

\subsubsection*{\textbf{Network Slice MANO}}
\label{sec:mano_systems:silces:mano}
A MANO system is supposed to orchestrate multiple complex management tasks in order to ensure the provisioning of network slice service. Thus, a MANO framework for 5G virtualized networks infrastructure is designed to go beyond providing the traditional Fault, Configuration, Accounting, Performance, and Security (FCAPS) management into providing additional management tasks. Some of the additional management functions, besides FCAPS are listed below:
\smallskip

 \begin{enumerate}
\item Software image management 
\item Service reliability management 
\item Policy management 
\item Bandwidth and Latency management
\item QoS/QoE management
\item Mobility management 
\item Energy management
\item Charging and billing management 
\item Network slice update/upgrade
\item VNF lifecycle management, including VNF scaling and migration 
\item Virtualized infrastructure management i.e., management of resource capacity, performance, fault, isolation etc.
\end{enumerate}
\smallskip

As mentioned earlier, the basic building block of a network slice at the virtualization layer is the VNF. The MANO entity performs the lifecycle management of a network slice by managing the individual VNFs that are part of the network slice.

\info{
\begin{itemize}
  \item Draft by: Michael (+Zarrar)
  \item \sout{Explain what a MANO system is and how it relates to NFV}
  \item Explain existing MANO systems, namely \sout{Tacker}, OSM, Open-O, OpenBaton, ECOMP, SONATA, NORMA
  \item Provide pointers to state-of-the-art research (papers), standards and organizations (e.g. ETSI, OPNFV), and open-source projects (Tacker, OSM, etc.). Maybe this could be a sub-section.
  \item Focus on SONATA~\cite{sonata.webpage} and explain the DevOps approach
\end{itemize}
}

\section{Software Defined Networking}
\label{sec:sdn}
\info{
\begin{itemize}
  \item Draft by: Fabian
  \item Explain SDN and focus on network programmability.
  \item Maybe: Avoid focusing on a particular technology, say OpenFlow
  \item Provide pointers to state-of-the-art research (papers), standards 
  and organizations (e.g. ONF), and open-source projects (ONOS, ODL, etc.). 
  Maybe this could be a recurring sub-section.
\end{itemize}
}
Evidently, introducing NFV and MANO systems into 5G networks also fosters very dynamic mechanisms of data traffic engineering and steering. For instance, new connections have to be set up fast and agile, e.g. to connect VNFs within and across data centers and establish end-to-end service. Likewise, network equipment has to be updated and (re-) configured continuously to support the NFV infrastructure and architecture. This, however, is very hard to do using traditional approaches for network operations where changes are often done (semi) manual at relatively long timescales, like minutes, hours, even days. To this end, Software Defined Networking comes into play to overcome the limitations of traditional networks and traditional network operations.

Software Defined Networking is a network para\-digm that evolved from work done at UC Berkeley and Stanford~\cite{mckeown:2008:01}. The motivation was to break up ossified networks by replacing rigid hardware-based proprietary equipment and services with deeply programmable common software-driven services and methods that span across multiple vendor-platforms; thereby decoupling the release cycles of agile software from the comparatively slow release cycles of integrated software and hardware. The radical change towards making networks programmable and enabling applications and network services to directly control the abstracted infrastructure, sparked a major development direction in research and education networks and commercial networks, affecting especially established network equipment vendors among the market players.

SDN has gained a lot of traction over the past years. The ability to manage network services through abstractions of lower level functionalities opens up a wide range of new architecture, management and operation options, including new forms of interaction between end-users’ applications and networks. Deploying agile software on white-box switches is expected to improve the cost-performance behavior of the network. Another great value of SDN will be the ability for rapid delivery of user services while using network resources more efficiently.

\subsection{OpenFlow and ONF}
\label{sec:sdn:onf}
An important protocol in the space of SDN is \textit{OpenFlow}, which enables the communication between network infrastructure elements and the network controlling, software-based entities. OpenFlow is maintained by the Open Networking Foundation (ONF) and today supported by all major network equipment vendors. 

From ONF's point of view, SDN has started as a vehicle to flexibly update packet forwarding algorithms. Since then, its applicability extended to the wider communications network domains covering all kinds of applications across the enterprise, carrier, data-center and campus network areas. Expanding from the initial three-layer architecture picture, consisting of Infrastructure, Control and Application Layers, the ONF published a detailed SDN Architecture~\cite{onf-arch-10, onf-arch-11} that will very briefly be introduced in the following. It is based on the following three principles:
\smallskip

\begin{itemize}
	\item Decoupling of control from traffic forwarding and processing. This is to enable independent deployment, life cycles and evolution of control and traffic forwarding and processing entities.
	\item Logically centralized control. Logically centralized means that control appears from the outside, application perspective as a single entity, but it is not implied to be deployed in a centralized monolithic implementation.
	\item Programmability of network services. Interfaces between SDN components expose resource abstractions and state. Applications are enabled to act on these abstractions  and states programmatically using a well defined API.
\end{itemize}
\smallskip

We believe that open interfaces and related protocols, like OpenFlow, are key for these systems that are built of decoupled functional components to enable the system operators to deploy components from any combination of a multitude of sources like commercial vendors and open-source groups. Open, well-defined interfaces may encourage competition between providers of community-agreed (standardized) functionality. However, proprietary features and interfaces should be expected to persist, especially in non-mainstream areas or for specialized ad-hoc extensions.

\begin{figure}
 \includegraphics[width=\linewidth]{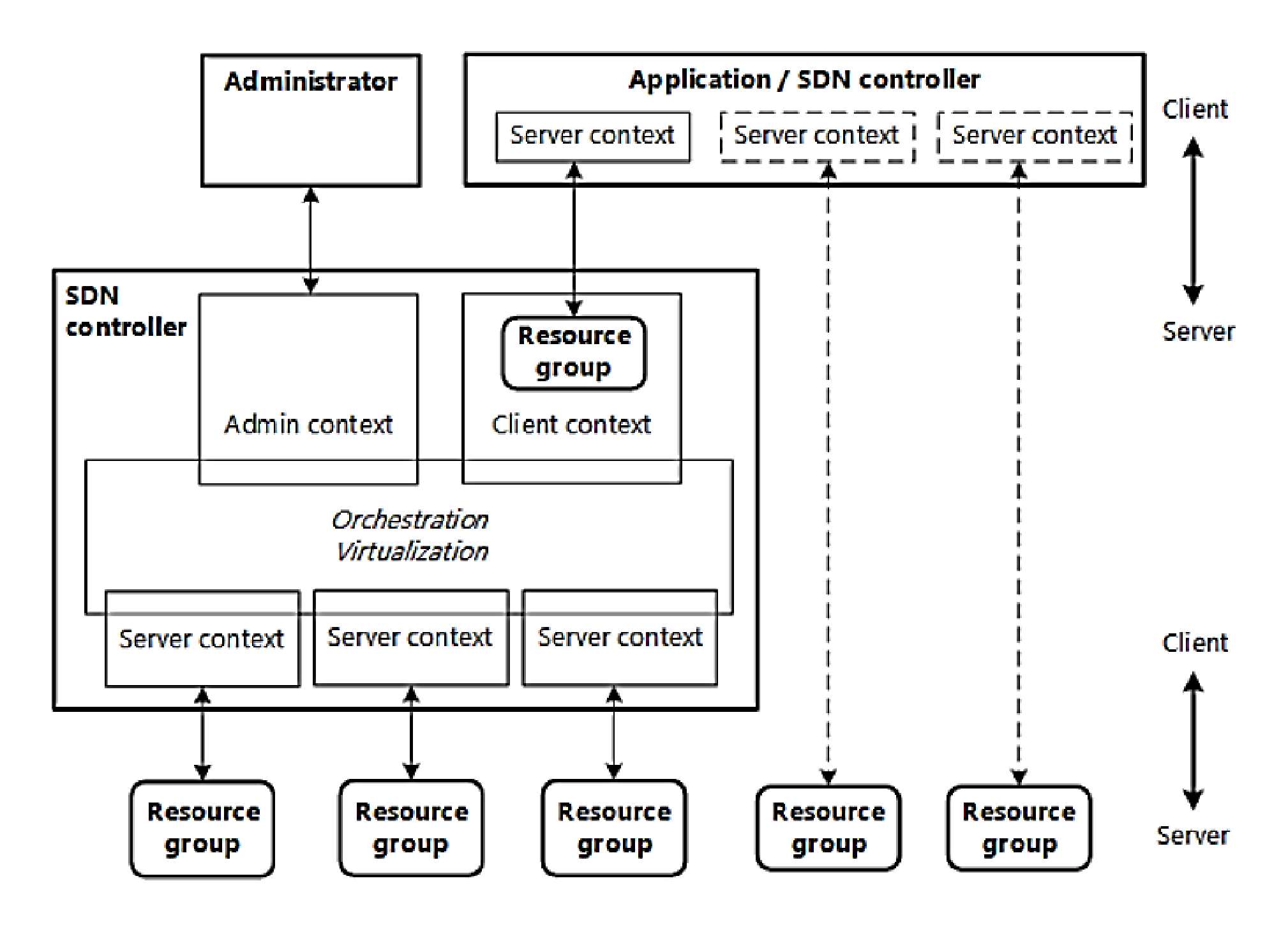}
 \caption{The ONF SDN architecture, adapted from~\cite{onf-arch-11}. It shows the SDN controller in the center of the architecture. It mediates between control applications at northbound interfaces and resources connected to the southbound interfaces.}
 \label{fig:sdn-architecture}
 \end{figure}

As shown in \myfigure{fig:sdn-architecture}, SDN controllers are at the center of the SDN architecture as envisioned by the ONF. They are responsible for the  provisioning, management and control of services and related resources. To this end, the controller offers so-called \textit{northbound interfaces} to applications and \textit{southbound interfaces} to the resources. Using these interfaces, users and applications have the ability to directly interact with the network. Leveraging the SDN controller’s northbound interface, authorized applications establish so-called management-control sessions in order to invoke services or to change the state of a resources at the southbound interface. In addition, the administrator role is responsible to create and maintain the environment needed to provide services to clients. It has the authority to configure the SDN controller, as well as to create and manage client and server contexts. To this end, configuring an SDN controller includes the creation of the controller itself, the installation and modification controller-internal policies, and the installation and configuration of actual resources and control applications.

Service consumption, i.e. data transfer and data processing, takes place through the corresponding network resources. Ultimately, user traffic is conveyed by physical resources, which may be any number of levels of abstraction below the resources visible to the client or to any particular SDN controller. Thus, leveraging the SDN controller's ability to abstract complex network resources and to mediate between resources and control applications, also paves the way to \textit{virtualized network resources}. As the controller manages the information flow from the resources to the applications, it can restrict the view and only provides a subset of resources or features to its upper layers. Thus, control applications can access network functionality and for example steer network traffic easily by using a standardized interface which simplifies dynamic network management a lot.




For a more detailed view on SDN we refer to~\cite{onf:2012:01} and references therein. Moreover, the authors in~\cite{schaller2017} provide a detailed description of the ONF SDN architecture and its relationships to other standardization efforts. A comprehensive survey of SDN can be found in~\cite{kreutz:2015:01}.

\subsection{SDN implementation}

SDN system developments are buzzing with new commercial and non-commercial network hardware, like switches and routers, and software, like SDN controllers. In the following we provide some examples out of the growing number of available options.

\subsubsection*{\textbf{SDN Hardware}}
At the advent of SDN dedicated silicon was not available. SDN switches were (and still are) implemented using either firmware that was translating SDN control protocol abstractions into the switching chips tables or by leveraging already programmable hardware such as Network Processing Units and FPGAs.  
At the time of writing first chips that natively support a programmable forwarding plane are available. Today, switches and other network equipment often support forwarding plane programmability by protocols like OpenFlow~\cite{openflow-spec-15}. Moreover, Ternary Content Addressable Memory (TCAM) in switches, which was often a bottleneck, is growing as well. This improves the practical applicability of programmable hardware. Bare-metal or white-box switches that do not ship with proprietary but allow various operating systems have arrived in the mass market. The market for operating systems to install on these switches is evolving rapidly and there is already a wide range of options available.

\subsubsection*{\textbf{SDN Software}}
Similar to SDN hardware, a diversity of SDN controllers has been developed meanwhile, too, and are available either on commercial basis or as open-source. Two very prominent examples are the OpenDaylight controller and the ONOS controller, both governed by the Linux Foundation.

\subsection{Future directions in SDN research}
The initial ideas that lead to the inception of SDN and OpenFlow came from 
the researchers desire to be able to innovate more freely and to be able to 
experiment with new Internet architectures. To some degree this goal 
is achieved by today's incarnations of SDN. We are now able to program the 
forwarding of packets in the network. For example this allows to explore 
and realize advanced and reactive traffic engineering approaches, it 
provides deep visibility into the current utilization of the network, or 
enable creative re-use of header fields in well known data plane protocols. 
\todo{add some citations}   

However contemporary implementations of SDN are focused on providing 
support for existing data-plane protocols and legacy approaches of 
operating networks. In particular, the match part of forwarding rules 
allows to specify subsets of network traffic based on well known header 
fields that are in use in today's networks. Likewise the action part allows 
to prescribe typical actions, such as output on port X, or drop. 

Limitations still exist when users want to define their own, new, data plane 
protocols, on which they want to match. Thus, enabling flexible matching on 
arbitrary header fields has been a recent topic of SDN research and 
standardization. Key works in this context are Protocol Independent 
Forwarding (PIF/P4)~\cite{p4-sigcomm14}, Protocol Oblivious Forwarding (POF)~\cite{pof-hotnets13} or Deeply Programmable Networks (DPN)~\cite{dpn}.

Extensions on the matching capabilities of SDN are especially interesting 
for new network architectures such as Information Centric Networking (ICN), 
Locator Identifier Split, and other approaches that define their 
own stacks of network protocols. Yet more recently, work on extending the 
action part of forwarding rules has caught the interest of the research 
community. The EU funded BeBa research project~\cite{beba.webpage} introduced two concepts that extend the programmability of 
actions. In a first contribution stateful flow processing has been added to SDN. Up to version 1.5, OpenFlow does not allow to store state from the processing of one packet of a flow (read match traffic identifier) to the next. OpenState~\cite{openstate-ccr14} is adding this possibility. In a second contribution, in-switch packet generation~\cite{insp-sosr16} adds the capability to programatically generate packets inside the switch, reacting to triggering packets or other in-switch 
events. 

These extension and future directions are particularly beneficial in the light of increased deployment of NFV. They allow to partially implement VNF functionality directly within the network elements, thereby reducing the need for deploying virtual machines.

\section{The marriage of SDN and NFV}
\label{sec:mariage_of_sdn_and_nfv}
\info{
\begin{itemize}
  \item ToDo: Brainstorming Zarrar, Fabian, Michael
  \item How does ONF see NFV?
  \item How does ETSI see SDN?
  \item How do open-source projects view SDN and NFV?
  \item No conclusive answers
\end{itemize}
}

\begin{figure}
 \includegraphics[width=\linewidth]{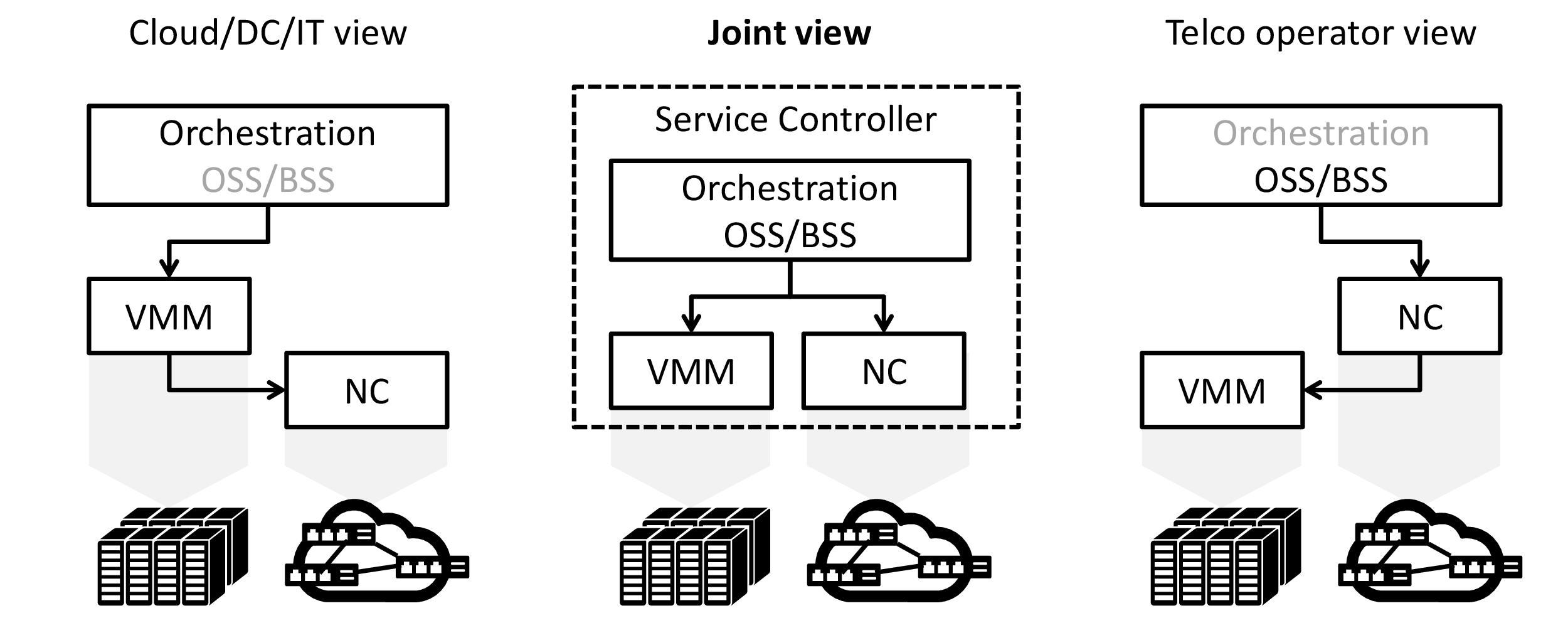}
 \caption{Different views of chains of command in the Cloud/DC/IT 
 	world (left) and the telecommunication world (right). A joint view is shown in the
 	middle, offering a way forward. 
 	(VMM: virtual machine management, NC: network controller)}
 \label{fig:marriage}
 \end{figure}
 
Bringing together SDN and NFV also means to bring together the views of so far distinct communities. On the one hand there is the cloud, data center and IT view that have long standing experience with deploying and managing virtual machines. And on the other hand there is telecom operator and vendor view, with a long lasting history of providing communication services. The problem that arises from these different views is illustrated in Fig.~\ref{fig:marriage}.

On the left side we show the typical chain of command in a typical data center deployment. The users and admins of the data center will use an orchestration interface to upload virtual machine images and request the instantiation of their service. In a first step the virtual machine management system, which is part of the Virtual Infrastructure Manager, will identify suitable servers and spin up the virtual machines according to the request. Then it will instruct a network controller component to provide connectivity between the instantiated virtual machines.

Contrary on the right side we show the typical telco view. Telcos main piece of infrastructure is the network which is controlled by the NC. Operation and business support systems are responsible for offering a unified central point for admins and customers to provision and monitor their services which mainly consist of providing connectivity typically with service level agreements. In these NFV times, several of the more complex network elements, which were previously managed by the network controller are now deployed as virtual machines. Thus the network controller will instruct the virtual machine management to instantiate a VNF. 

To summarize, in both cases the network controller or the virtual machine manager are just seen to provide a service to the other. In order to bring these two worlds together we need to put them on the same level and integrate them further. This is shown in the middle and already implemented in SONATA. \todo{@Michael: Check!} The virtual machine manager and the network controller will become components of a single infrastructure resource controller. This is already supported by the ONF SDN architecture~\cite{onf-rel-sdn-nfv}, when assuming that the infrastructure can include compute and storage resources beyond the typically discussed network resources.

In order to fully utilize such a combined \emph{service controller}, we need to start developing holistic service descriptions for Internet applications, that include all its components, supported deployment topology, hints on how to scale, requirements on network path properties, desired network functions, as well as easy to program interfaces that abstract away unnecessary complexity from the developer.  

As described in the previous section, ONF takes a much broader view of network systems, and thus the broad definition of SDN that has developed over time within the ONF can be translated into many different ways in terms of specifications and implementations. ETSI NFV, on the other hand, provides a very precise architectural framework for a very clear purpose, and that is to manage and orchestrate NFV Infrastructure resources, typically located in data centers, that are utilized and consumed by telco related functions and services. In this context ETSI NFV specifies features and functions it requires from SDN. They then look into various possibilities of positioning SDN in the larger scope of NFV. 
From this perspective, the ETSI NFV system as per today's requirements uses the services of SDN to provide a programmable platform for establishing links between VNFs and VNF components, and to support enhanced functions such as policy based management of traffic between VNFs, or dynamic bandwidth management. Thus the NFV system realizes a fully programmable end-to-end network services within the NFV domain.

\begin{figure}[h!]
 \includegraphics[trim={2cm 1.5cm 3cm 1.5cm}, clip, width=\columnwidth]{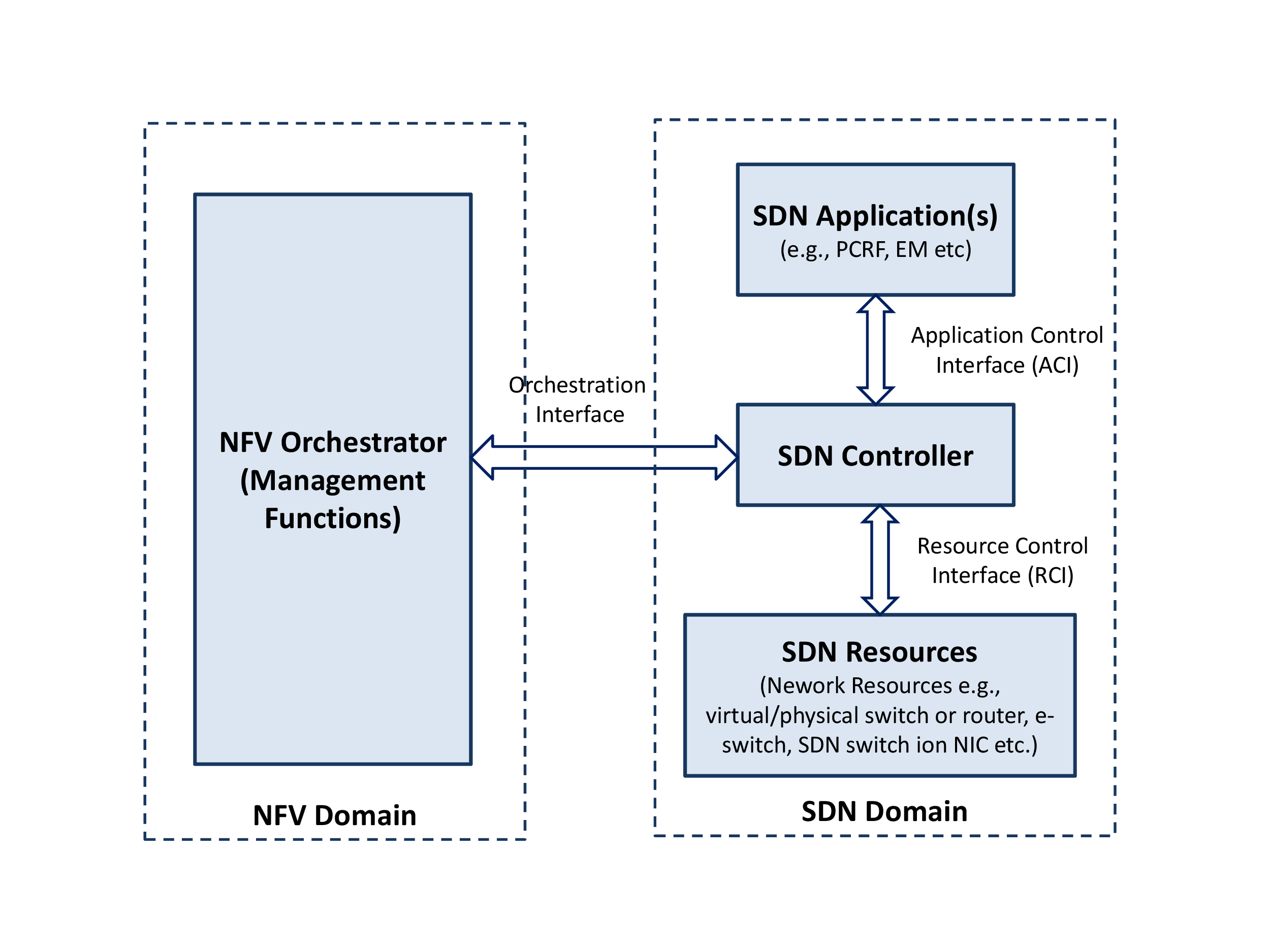}
 \caption{ETSI NFV perspective of interacing with the SDN domain \cite{etsi-eve}}
 \label{fig:sdnNfvGeneral}
 \end{figure}

When integrating the SDN functional components within the NFV infrastructure, it must take into consideration the SDN interfaces relevant for its requirements. Figure \ref{fig:sdnNfvGeneral} gives a high level overview depicting ETSI NFV perspective on interfacing with the SDN domain \cite{etsi-eve}. As shown, ETSI NFV is in the process of specifying the \emph{orchestration interface(s)} for interfacing the SDN controller with the NFV MANO system. These specifications take the interfaces internal to the SDN domain into account. That is, the Application Control Interface that provides to the VNFs an application programmatic control of abstracted network resources \cite{etsi-eve}, and the Resource Control Interface for controlling the NFV Infrastructure network resources (e.g, physical/virtual routers and switches, and networks connecting VNFs).

\begin{figure}[h!]
 \includegraphics[trim={0cm 0cm 0cm 1cm}, width=0.48\textwidth]{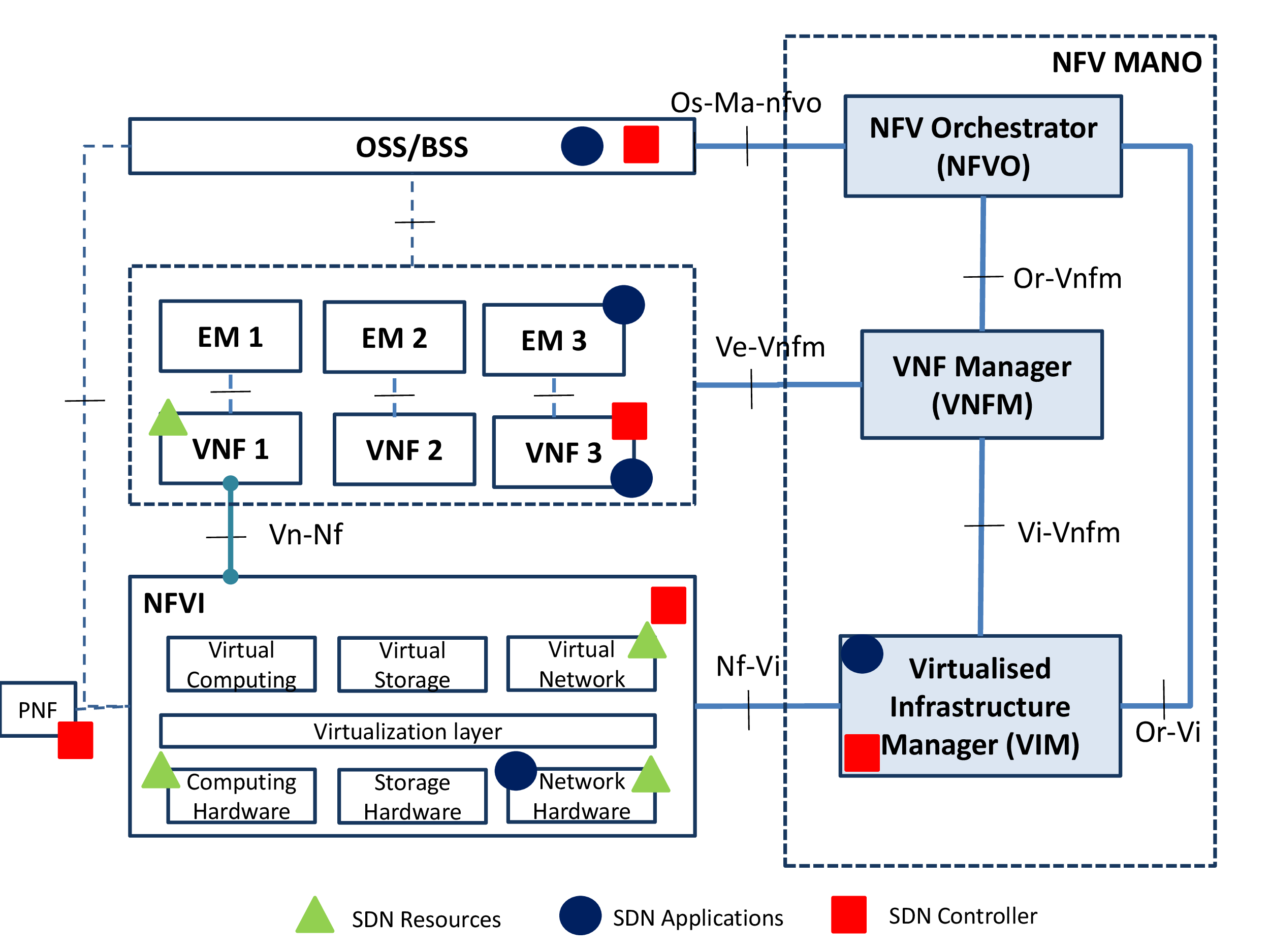}
 \caption{Possible options of positioning SDN Resources, SDN Controller and SDN Applications in NFV Architectural Framework}
 \label{fig:sdnNFV}
 \end{figure}

 In this context, ETSI NFV has published a detailed report \cite{etsi-eve} describing the various possible options of SDN federation in NFV. Figure \ref{fig:sdnNFV} summarizes these possible options of integrating \emph{SDN application}, \emph{SDN resources} and \emph{SDN controller} with different entities within the NFV MANO and NFV architecture. Each one has its own requirements on the NFV MANO interfaces. For example, there are five integration options for SDN controller to either (i) be part of OSS/BSS, (ii) exist as an entity within the NFV Infrastructure, (iii) exist as a Physical Network Function, (iv) be instantiated as a VNF, or (v) be integrated within the Virtual Infrastructure Manager. The latter approach is supported by the ONF SDN architecture \cite{onf-rel-sdn-nfv} and is also adopted by open source OPNFV platform \cite{opnfv.webpage}, where SDN controllers like ODL and ONOS are integrated with OpenStack, the latter being widely accepted as a suitable virtual machine management platform. The goal of OPNFV project is to provide a carrier grade integrated  open source reference platform for NFV. In other words, it is an ongoing project attempting the marriage between NFV and SDN. There are also some prominent research projects like 5G NORMA \cite{norma.webpage} that leverages on the SDN and NFV concepts in order to develop a novel mobile network architecture that shall provide the necessary adaptability in a resource efficient way able to handle fluctuations in traffic demand resulting from heterogeneous and dynamically changing service portfolios and to changing local context. From the NFV perspective 5G NORMA extends the NFV MANO framework to support multi-tenancy and manage service slices that may be extended over multiple sites. From the SDN perspective, it defines two SDN-based controllers, one for the management of network functions local to a mobile network service slice, and the second for the management of network functions that are common/shared  between mobile network service slices \cite{normaD32}. These controllers leverages on the concept of SDN controller and translate decisions of the control applications into commands to VNFs. 5G NORMA recommends these special SDN controllers to be deployed as VNFs. 

Thus, Figure \ref{fig:sdnNFV} gives different options of integrating the SDN system (application, resources and controller) in the context of NFV and \cite{etsi-eve} provides an overview of each option and its combination. The key point is that NFV aims at leveraging the programability feature of SDN in order to implement network services that may be designed according to some pre-configured VNF Forwarding Graph, or implement NS that may require the chaining of VNFs based on some policy/service or even based on VNF processing, for example, a security related VNF may want to change the path of traffic on the fly depending on its processing output. 

\begin{figure}[h!]
 \includegraphics[trim={0cm 0cm 0cm 0cm}, width=0.48\textwidth]{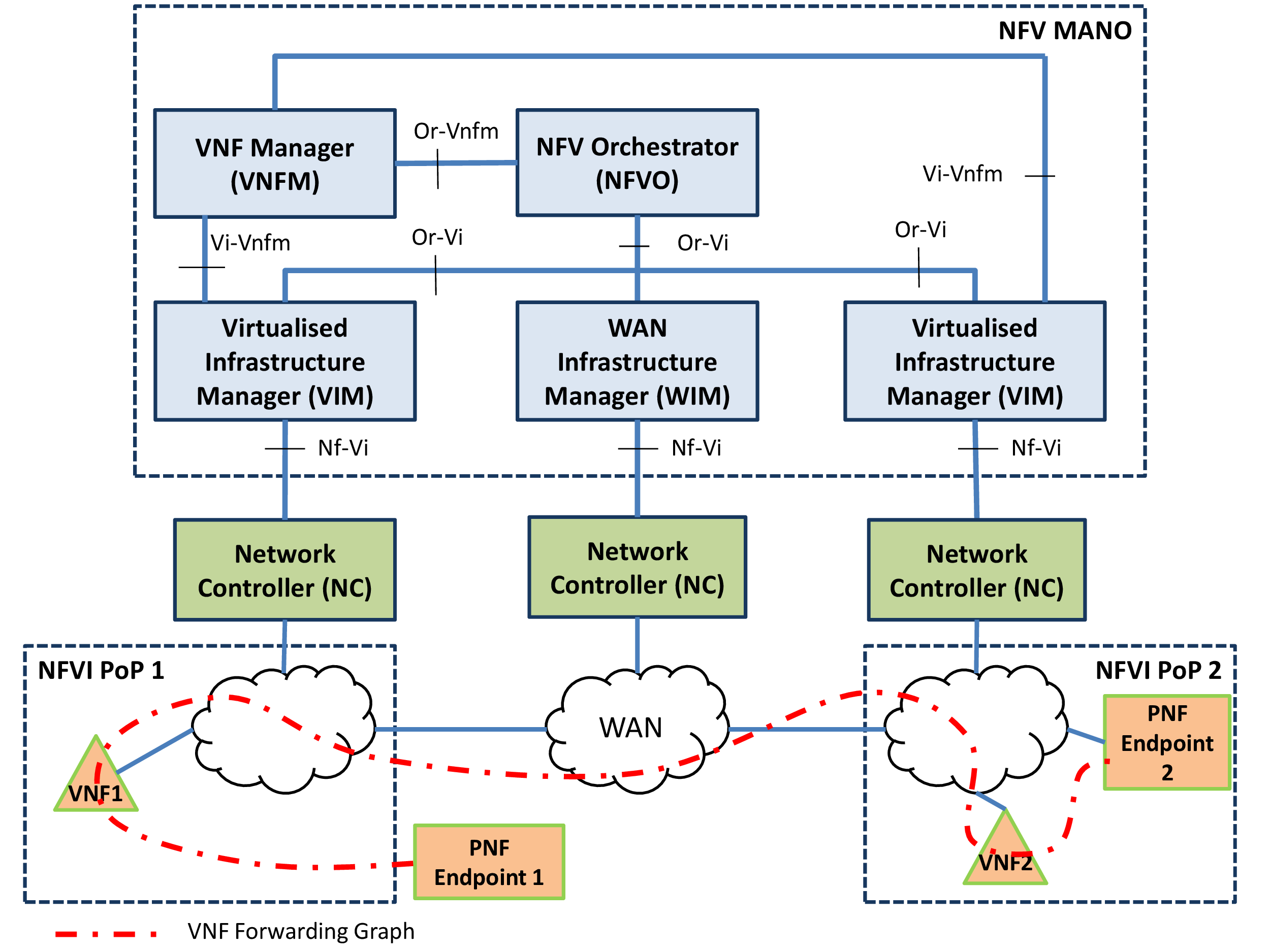}
 \caption{NFV SDN in a multi-site scenario}
 \label{fig:multiSite}
 \end{figure}

ETSI MANO in \cite{etsi-mano} provides clear insight as to how it can utilize the features of SDN for its respective purposes. Figure \ref{fig:multiSite} gives a useful overview with reference to a multi-site scenario where two network services involving two virtual (i.e., VNF1 and VNF2) and two physical network functions (i.e., PNF1 and PNF2) are extended over two NFV Infrastructure Points-of-Presence (PoP). Each NFVI-PoP has its own VIM while a WAN Infrastructure Manager (WIM) is also required for requesting connectivity services between the two NFVI-PoPs over the WAN. Multiple connectivity services are requested by the NFV Orchestrator over the Or-Vi interface from the respective VIMs/WIMs for establishing connectivity within their respective domains. Each VIM/WIM can request for the provisioning of virtual networks from the Network Controller (NC) over a fully open and programmable Nf-Vi interface. The NC, which for all practical purposes can be an SDN controller and will be referred to as such. This SDN controller has visibility into the devices (i.e, SDN resources) that they control directly and thus is able to provide an abstracted view of them to the VIM/WIM via the Nf-Vi northbound interface. It should be noted that the SDN applications can also reside inside VIM (see Figure \ref{fig:sdnNFV}). The SDN controller then establishes the connectivity services by configuring the forwarding tables of the underlying VNFs/PNFs. Although shown as a separate functional entity, the SDN controller can also be part of VIM/WIM as discussed above (see Figure \ref{fig:sdnNFV}). 
At the time of writing this paper, the Infrastructure and Architecture (IFA) working group of the ETSI ISG for NFV is specifying use cases for multi-site connectivity in order to draw more concrete requirements for the Northbound interface (i.e., Nf-Vi) of the SDN controller in order to achieve a happy successful marriage between NFV and SDN.

%

\section{Conclusion}
\label{sec:conclusion}
%

Communication networks are currently undergoing a major evolutionary change in order to be capable to flexibly serve the needs and requirements of massive numbers of connected users and devices and to enable the functioning of the new set of envisioned applications and services in an agile and programmable way. Key terms in that context are Internet-of-things, virtualization, softwarization and cloud-native.

In order to be able to maintain and run these networks over 5G slices, NFV and SDN technologies are widely considered as the key enablers in network architecture, design, operation and management. Several organizations (ETSI NFV, ONF, ETSI MEC, NGMN, 3GPP, IEEE, BBF, MEF etc.) are working on standardizing the architecture frameworks and interfaces required for combining the multitude of components into a functional system that can be implemented within the provider/operator systems based on a variety of business models and use cases. 

In parallel to standardization activities, several components are  being developed under the umbrella of open-source projects (OpenStack, OPNFV, OSM, ONAP, ODL, ONOS, etc.) that are expected to complement, if not replace, commercial vendors' products. Moreover, these open-source projects and relevant standardization bodies are also mutually influencing each other towards the development of their respective goals, and validating and progressing their respective work.

Although the community/industry is on its way and progressing well to realize a large part of the 5G vision by 2020, a number of research challenges and issues are still open that needs to be addressed in order to ensure a healthy conception of the envisaged 5G systems. Some of those questions/issues are:

\begin{itemize}
  \item How to manage/handle the agility of software, especially in view of the trend/need of the decomposition of network function into micro-functions.
  \item How to distribute network functions onto different  execution platforms, when highly programmable hardware (switches, smartNICs) becomes more ubiquitous.
	\item How to ensure interoperability between different vendors, especially in this cloud-native environment of massively decomposed network functions.
	\item What is the best way of translating and mapping the customers/clients/tenants business requirements over the service providers' infrastructure. 
	\item How to ensure that the QoS and QoE requirements and SLAs can be fulfilled in the cloud-native environment.
	\item How to efficiently assign and manage resources for the multitude of slices that may exist within the same administrative domain, or traverse over different administrative domains.
	\item How, and to what extent  can the network/system management be automated in order to reduce the need for manual tasks and intervention.
\end{itemize}

\section*{Acknowledgment}
\noindent\small{This research work has been performed in the framework of H2020-ICT-2014-2 project 5G NORMA and SONATA projects. The authors would like to acknowledge the contributions of their colleagues of the 5G NORMA and SONATA partner consortium, although the views expressed are those of the authors and do not necessarily represent the project.}



\bibliographystyle{IEEEtran}
\bibliography{IEEEabrv,myBib}
\end{document}